\documentclass[conference]{IEEEtran}
\IEEEoverridecommandlockouts
\usepackage{cite}
\usepackage{amsmath,amssymb,amsfonts}
\usepackage{graphicx}
\usepackage{textcomp}
\usepackage{xcolor}
\def\BibTeX{{\rm B\kern-.05em{\sc i\kern-.025em b}\kern-.08em
    T\kern-.1667em\lower.7ex\hbox{E}\kern-.125emX}}

\usepackage{amsmath,amssymb,amsfonts}
\usepackage{graphicx}
\usepackage{textcomp}
\usepackage{xcolor}
\usepackage{cases}
\usepackage{algorithm}
\usepackage[noend]{algpseudocode}
\usepackage{tikz} 
\usepackage{cite}
\usepackage{comment}
\usepackage{balance}
\usepackage{subfigure}
\usepackage{color}

\usepackage[thmmarks, amsmath, thref]{ntheorem}

\theoremstyle{plain}
\theoremheaderfont{\upshape\bfseries}
\theoremseparator{.}
\theorembodyfont{\upshape}
\newtheorem{theorem}{Theorem}

\newtheorem{assumption}{Assumption}
\newtheorem{definition}{Definition}
\newtheorem{remark}{Remark}
\newtheorem{proposition}{Proposition}

\theoremstyle{nonumberplain}
\theoremseparator{:}
\theorembodyfont{\upshape}
\theoremsymbol{\ensuremath{\color{black}\blacksquare}}

\usepackage{caption}
\usepackage{yhmath}
\usepackage{enumitem}

\newcommand\independent{\protect\mathpalette{\protect\independenT}{\perp}}
\def\independenT#1#2{\mathrel{\rlap{$#1#2$}\mkern2mu{#1#2}}}

\renewcommand{\thepage}{\arabic{page}}
\renewcommand{\thesection}{\arabic{section}}
\renewcommand{\theequation}{\arabic{equation}}
\renewcommand{\thefigure}{\arabic{figure}}
\renewcommand{\thetable}{\arabic{table}}

\allowdisplaybreaks

\usepackage{xspace}
\usepackage{bbm}
\usepackage{mathrsfs}

\newcommand{\Pcal}{\mathcal{P}}

 \def\E{\mathsf{E}}

\def\textiid{i.i.d.\@\xspace}
\newcommand\iid{\ifmmode\text{ i.i.d. } \else \textiid \fi}

\newcommand{\Real}{\mathbb{R}}

\makeatletter
\def\thickhline{%
  \noalign{\ifnum0=`}\fi\hrule \@height \thickarrayrulewidth \futurelet
   \reserved@a\@xthickhline}
\def\@xthickhline{\ifx\reserved@a\thickhline
               \vskip\doublerulesep
               \vskip-\thickarrayrulewidth
             \fi
      \ifnum0=`{\fi}}
\makeatother

\newlength{\thickarrayrulewidth}
\title{Identifying Direct Causes\\ using Intervened Target Variable}

\begin{document}

\author{\IEEEauthorblockN{Kang Du and Yu Xiang}
\IEEEauthorblockA{
Department of  Electrical and Computer Engineering\\
University of Utah\\
\{kang.du, yu.xiang\}@utah.edu
}

\and

\IEEEauthorblockN{Ilya Soloveychik}
\IEEEauthorblockA{Department of Statistics\\
Hebrew University of Jerusalem\\
soloveychik.ilya@mail.huji.ac.il} 
}

\maketitle

\begin{abstract}
Identifying the direct causes or causal parents of a target variable is crucial for scientific discovery. Focusing on linear models, the invariant prediction framework was built upon the invariance principle, namely, the conditional distribution of the target variable given its causal parents is invariant across multiple environments or experimental conditions. However, their identifiability results for causal parents can be restrictive with respect to the underlying graph structure and the experimental conditions for generating interventional data. Motivated by a recent alternative formulation of invariance, called the invariant matching property, we establish identifiability results under relatively mild assumptions, which leads to a simple yet effective procedure for identifying causal parents. We demonstrate the performance of the proposed method over various synthetic and real datasets.

\end{abstract}

\begin{IEEEkeywords}
Causal parents, linear models, invariance, identification.
\end{IEEEkeywords}

\section{Introduction}

\noindent The problem of identifying the ``direct causes" (or causal parents) of a target variable $Y\in \Real$ from a vector of correlated features $X = (X_{1},\ldots,X_{d})^{\top} \in \mathcal{X} \subseteq\Real^d$ is a fundamental yet very challenging question. When all the data is collected from one environment, perhaps the most related work is the LiNGAM method~\cite{shimizu2006linear} for the discovery of linear causal models under the non-Gaussian noise assumption. Datasets are often available from multiple environments, which has attracted much attention in recent years, where observations of $(X^{e},Y^{e})$ are collected from a set of environments (e.g., different experimental settings) denoted by $\{e \in \mathcal{E}\}$. In this setting, the multiple environments can be leveraged through the lens of invariance, which leads to the invariant causal prediction (ICP) method for identifying causal parents~\cite{peters2016causal}; the critical assumption is called the invariance principle, stating that there exists $S \subseteq \{1,\ldots,d\}$ such that
\begin{equation}
    \Pcal_e(Y|X_S) =  \Pcal_h(Y|X_S),
    \label{eq:ICP}
\end{equation}
for all $e,h \in \mathcal{E}$, where $X_S$ denotes the set of features in set $S$. The ICP method is focused on linear models and it has been extended to various settings (see~\cite{heinze2018invariant,pfister2019invariant,gamella2020active} and references therein) along with numerous applications (e.g.,\cite{meinshausen2016methods,goddard2022invariance}). However, this invariance principle breaks down when $Y$ is intervened, which is likely to take place in real-world applications, considering that there are various ways to intervene on $Y$ in linear models.  

One natural question is \emph{whether an alternate form of invariance exists in this setting}. One attempt called the invariant matching property (IMP) method has been made in~\cite{du2022learning} to address this under the assumption that at least one child is not intervened (with a computation-efficient version when only $Y$ is intervened in~\cite{du2023generalized}). To shed light on this problem, motivated by the IMP method, we focus on the \emph{multiple-environment} setting and provide the identifiability of parents of $Y$ (denoted by $PA(Y)$) in Section~\ref{sec:ident} and we propose a simple yet effective voting procedure to estimate $PA(Y)$ in finite samples in Section~\ref{sec:alg}. We evaluate the proposed method using both synthetic and real-world datasets. 

\section{Background and Problem Formulation}
\subsection{Invariant Causal Prediction (ICP)~\cite{peters2016causal}}
\label{sec:ICP}

We start by briefly describing the ICP which is focused on linear structural causal models (SCMs) in~\cite{peters2016causal}. Consider $(X^{e},Y^{e})$, $e \in \mathcal{E}$, generated by linear SCMs
\begin{numcases}
{\mathcal{M}^{e}:}
    X^{e} = \alpha^{e} Y^{e} + B^{e} X^{e}+ \varepsilon_{X}^{e}  \nonumber\\
    Y^{e} = (\beta^{e})^{\top}X^{e} +  \varepsilon_{Y}^{e}, \nonumber
\end{numcases}
where $\varepsilon_{X}^{e} = (\varepsilon_{X,1}^{e},\ldots,\varepsilon_{X,d}^{e})^{\top}$ and $\varepsilon_{Y}^{e}$ are independent noise variables. The causal graph $\mathcal{G}(\mathcal{M})$ induced by $\mathcal{M}^{e}$ can be drawn according to the non-zero entries in $\{\gamma^{e},B^{e},\beta^{e} \}$; the connectivity of the graph does not change with the environment $e$ (i.e., $\mathcal{G}(\mathcal{M})=\mathcal{G}(\mathcal{M}^e)$ for all $e$). The change in parameters in the assignment of a variable (e.g., its coefficients and noise variables) captures the change in experimental conditions, which is referred to as \emph{intervention on that variable}. One special case of the varying $\beta^e$ has been studied in~\cite{du2020causal}.

Under linear SCMs, the invariance described in~\eqref{eq:ICP} holds for $S= PA(Y)$ when both $\beta^{e}$ and the distribution of $\varepsilon_{Y}^{e}$ are invariant across environments, for which we call \emph{$Y$ is not intervened}. Under some sufficient conditions (see~\cite[Theorems~2 and~3]{peters2016causal}), the identification of $PA(Y)$ can be established by assuming the existence of $Y^{e} = b^{\top}X_{S}^{e} + \varepsilon_{Y}^e$ with $X_{S}^{e} \independent \varepsilon_{Y}^e$
for some $b \in \mathbb{R}^{|S|}$ and Gaussian noise that is invariant ($\varepsilon_{Y}^{e}\sim \varepsilon_{Y}$). Roughly speaking, the sufficient conditions in~\cite[Theorems~2 and~3]{peters2016causal} require either every $X_{j}$ to be intervened once or a special class of graph structures. In practical experiment settings, however, the conditions mentioned above can lead to two potential issues: (1) the special graph structure may not hold, and (2) intervening on every $X_{j}$ (while keeping $Y$ to be not intervened) can be costly. Our approach in this work is partially motivated by the following observation: It is possible to relax the two restrictive conditions by considering the opposite setting, i.e., only $Y$ is intervened,
\begin{numcases}
{\mathcal{M}^{1,e}:}
    X^{e} = \alpha Y^{e} + B X^{e}+ \varepsilon_{X}^{e}, \nonumber \\
    Y^{e} = (\beta^{e})^{\top}X^{e} +  \varepsilon_{Y}^{e}, 
  \nonumber 
\end{numcases}
where we consider the setting when the noise distributions are invariant ($\varepsilon_{Y}^{e}\sim \varepsilon_{Y}$ and $\varepsilon_{X}^{e}\sim \varepsilon_{X}$). Under $\mathcal{M}^{1,e}$, the invariance principle~\eqref{eq:ICP} no longer holds in general, and, in particular, it fails to hold for $S= PA(Y)$.

\subsection{Invariant Matching Property (IMP)~\cite{du2022learning}}

The invariant matching property was introduced for the prediction of $Y$ in unseen environments. In particular, it handles interventions on both $X$ and $Y$. Its definition relies on linear minimum mean squared error estimators, which we define as follows.  
 For a target variable $Y \in \mathbb{R}$ and a vector of predictors $X\in \mathbb{R}^{p}$ with a joint distribution $\Pcal$, let $\theta^{\text{ols}}$ denotes the population ordinary least squares (OLS) estimator, then the LMMSE estimator is defined as 
\begin{equation*}
    \E_{l,\Pcal}[Y|X] := (\theta^{\text{ols}})^{\top}(X-\E[X])+\E[Y].
\end{equation*}  

\begin{definition}[\!\!\cite{du2022learning}]
\label{def:imp}
For $k \in \{1,\ldots,d\}$, $R \subseteq \{1,\ldots,d\}\setminus k$, and $S \subseteq \{1,\ldots,d\}$, we say that the tuple $(k,R,S)$ satisfies the \emph{invariant matching property} (IMP) if, for every $e \in \mathcal{E}$, 
 \begin{align}
    \E_{l, \Pcal_e}[Y\,|\,X_{S}] 
    = \lambda \E_{l, \Pcal_e}[X_{k}\,|\,X_{R}] + \eta^{\top}X^{e}  \label{invar_coeff} ,
 \end{align}
 for some $\lambda \in \mathbb{R} $ and $\eta \in \mathbb{R}^{d}$ that do not depend on $e$. 
\end{definition}

The IMP holds when $ \E_{l, \Pcal_e}[Y\,|\,X_{S}] $ can be represented as an invariant linear function of $X$ together with an additional feature $\E_{l, \Pcal_e}[X_{k}\,|\,X_{R}]$. The term ``matching" comes from the following observation: the way that the coefficients of $ \E_{l, \Pcal_e}[Y\,|\,X_{S}] $ change across environments should match the changes of the coefficients of $\E_{l, \Pcal_e}[X_{k}\,|\,X_{R}]$, thus a constant scaling parameter $\lambda$ will cancel out the changes and result in $\eta^{\top}X^{e}$ with invariant coefficients. The idea of IMP is illustrated via toy examples in~\cite[Section~$3$]{du2022learning}. In practice, the linear relation can be estimated from the training data and reused on the test data, while the additional feature can be estimated from the unlabeled data in every environment. However, the invariant coefficients in the IMP are not identifiable when multicollinearity occurs, thus the conditions in the proposition below are necessary.

\begin{proposition}[\!\!\cite{du2022learning}]\label{prop:unique_theta}
For a tuple $(k,R,S)$ that satisfies an IMP, the  parameters $\{\lambda,\eta\}$ can be uniquely identified in $\mathcal{E}$ if $|\mathcal{E}|\geq 2$ and
\begin{equation}
    \E_{l, \Pcal_e}[X_{k}|X_{R}=x] \neq \E_{l, \Pcal_h}[X_{k}|X_{R}=x] 
    \label{eq_assup_unique}
\end{equation}
for some $e,h \in \mathcal{E}$ and $x \in \mathcal{X}_{R}$.  
\end{proposition}
An example that violates~\eqref{eq_assup_unique} is when $X_{k}$ and $X_{R}$ are independent and $X_{k}$ has an invariant mean. In this case, the IMP holds when $Y$ and $X_{S}$ are also independent and $Y$ has an invariant mean. To avoid such corner cases, we mainly focus on IMPs such that $\lambda$ is identifiable (i.e.,~\eqref{eq_assup_unique} holds). Interestingly, certain IMPs will imply an alternate form of invariance, 
\begin{equation}
    \Pcal_e(Y|\phi_{e}(X)) =  \Pcal_h(Y|\phi_{h}(X)),\label{inva_trans}
\end{equation}
where $\phi_{e}(X^{e}) := (X_{S}^{e},\E_{l,\Pcal_e}[X_{k}|X_{R}])^{\top}$, which includes the invariance principle~\eqref{eq:ICP} as a special case.

Several classes of IMPs have been characterized in~\cite{du2022learning}, while we focus on the setting when only $Y$ is intervened (due to the fact that interventional data is less costly to obtain in this case). For making predictions, the set $X_{S}$ should include as many as $X_{j}$'s since the prediction error of the IMP is reduced when more predictors are included. As a result, $PA(Y)$ should naturally be included as a subset of $S$. The following proposition is a special case of~\cite[Theorem~$1$]{du2022learning}, showing that all $R$ and $S$ that contain $PA(Y)$ can be employed to construct IMPs.

\begin{assumption}\label{assump1}
$Y$ has at least one child in $\mathcal{G}(\mathcal{M}^{1})$. 
\end{assumption}

\begin{proposition}[\!\!\cite{du2022learning}]\label{prop:exists}
Under Assumption~\ref{assump1}, and model $\mathcal{M}^{1,e}$ and condition~\eqref{eq_assup_unique}, the IMP holds for any $(k,R,S)$ such that $k \not\in PA(Y)$ and $ PA(Y) \subseteq S, R $. 
\end{proposition}

Since our goal is to identify $PA(Y)$, the sufficient conditions for IMPs to hold in Proposition~\ref{prop:exists} are not directly useful. In the next section, we present the   
necessary conditions for the IMP that \emph{almost} overlap with the sufficient conditions. 


\section{Identifiability}
\label{sec:ident}

Let $\mathcal{I} = \{(k_{i},R_{i},S_{i})\}$ denote the set of all IMPs with $\lambda \neq 0$. Our identifiability results rely on the idea that
\begin{equation}
    PA(Y) = \cap_{i \in \mathcal{I} }R_{i}, \nonumber
\end{equation}
which says that $PA(Y)$ is included in $R$ for every IMP. 
 It turns out the sufficient condition for this is rather simple, i.e., none of the coefficients of the parents of $Y$ is invariant, namely, $\beta^{e}$ satisfies the following condition:
\begin{equation}
    \text{$  \beta_{j}^{e} \neq \beta_{j}^{h}$  for every $j \in PA(Y)$, for some $e,h \in \mathcal{E}$}.
    \label{eq:non_inv}
\end{equation}
In other words, the interventions on $Y$ are strong enough to influence the relation between $Y$ and each parent of $Y$. We provide the formal identifiability results below for an important class of IMPs from $\mathcal{I}$. 

\begin{theorem}\label{thm_ident}
Under Assumption~\ref{assump1} and model $\mathcal{M}^{1,e}$ such that~\eqref{eq:non_inv} holds, the IMP holds for $(k,R,S)$ with $PA(Y) \subseteq S$ and an identifiable $\lambda\neq 0$, only if $PA(Y) \subseteq R$. 
\end{theorem}

\begin{IEEEproof}
First, given $ PA(Y) \subseteq S $, the proof of~\cite[Theorem~$1$]{du2022learning} implies that there exist $c \in \mathbb{R}$ and $\alpha \in \mathbb{R}^{d}$ such that 
\begin{equation}
    \E_{l, \Pcal_e}[Y\,|\,X_S] = (c\beta^{e} +\alpha)^{\top}X^{e} \nonumber.
\end{equation}
Since the assumption that the nonzero parameter $\lambda $ is identifiable implies that
\begin{equation}
    \E_{l, \Pcal_e}[X_{k}|X_{R}=x] \neq \E_{l, \Pcal_h}[X_{k}|X_{R}=x] \nonumber
\end{equation}
for some $e,f \in \mathcal{E}$ and $x \in \mathcal{X}_{R}$, the coefficients of  $\E_{l, \Pcal_e}[Y|X_{S}]$ can not be all invariant. Thus we have $c\neq 0 $. 

Now we proceed to prove the theorem by contradiction. If $PA(Y) \subseteq S$,
consider $R \subseteq \{1,\ldots,d\}$ such that $l \not \in R$ for some $l \in PA(Y)$, then 
\begin{equation}
    \E_{l, \Pcal_e}[X_{k}\,|\,X_{R}] = \sum_{j\in R} \gamma_{j}^{e}X_{j}^{e}, \nonumber
\end{equation}
for some coefficients $\{\gamma_{j}^{e}\}$.
Then the IMP holds with parameters $\lambda$ and $\eta$ only if 
\begin{equation}
(c\beta^{e} +\alpha)^{\top}X^{e} - \lambda \sum_{j\in R} \gamma_{j}^{e}X_{j}^{e} -\eta^{\top}X^{e} = 0. \nonumber
\end{equation}
However, the coefficient of $X_{l}$ is $c\beta_{l}^{e}+\alpha_{l}-\eta_{l}$ for each $e \in \mathcal{E}$, which cannot be zero for all environments due the assumption on $\beta^{e}$, which yields a contradiction. 
\end{IEEEproof}
\smallskip

The restrictions on the class of IMPs can be relaxed. For instance, $\lambda \neq 0$ can be removed since it holds with probability one if the variance of $\varepsilon_{Y}$ is sampled from a continuous distribution. The proof relies on the explicit calculations of $\E_{l, \Pcal_e}[Y\,|\,X_S] $ using~\cite[Lemma~$1$]{du2022learning}. To satisfy the condition that $PA(Y) \subseteq S$, we focus on IMPs that are more predictive for $Y$, which is generally achieved by the two procedures proposed in~\cite{du2022learning}.

\begin{remark}
One may wonder if an alternate idea using $PA(Y) = \cap_{i \in \mathcal{I} }S_{i}$ would also work. This is, however, not the case. Since the IMP was originally proposed for the prediction of $Y$, IMPs with $S_{i}=PA(Y)$ may be discarded by the procedures proposed in~\cite{du2022learning}, as the prediction error can be further reduced when some children of $Y$ are included in $S$. But focusing $R$ allows us to preserve the IMPs with $R_{i} = PA(Y)$.
\end{remark}
\smallskip

\section{Algorithm}\label{sec:alg}

 Two procedures, called  $\text{IMP}$ and $\text{IMP}_{\text{inv}}$, have been proposed in~\cite{du2022learning} for identifying IMPs from data. The former procedure is derived from the definition of IMP and the latter is based on the invariance implied by IMP (i.e.,~\eqref{inva_trans}). Due to page limit, we refer the readers to ~\cite[Section~$5$]{du2022learning} for a full description of both procedures. In this section, we propose a simple voting procedure to identify $PA(Y)$ using estimated IMPs obtained from the two procedures.

Given \iid samples of $(X^{e},Y^{e})$ from a set of environments $\{e \in \mathcal{E}\}$. The two procedures are designed to test 
\begin{equation}
    \mathcal{H}_{0}: (k,R,S) \text{ satisfies an IMP } 
\end{equation}
for every $(k,R,S)$, $k \in \{1,\ldots,d\}$, $S \subseteq \{1,\ldots,d\}$, and $R \subseteq S \setminus k$. Additionally, a prediction score was employed in~\cite{du2022learning} to further select IMPs that are more predictive for $Y$. Let $\hat{\mathcal{I}} = \{(k_{i},R_{i},S_{i})\}_{i=1}^{q}$ denote the set of identified IMPs. Motivated by the identifiability results in Section~\ref{sec:ident}, we propose a voting procedure to estimate $PA(Y)$ as follows. Let $V^{(0)} \in \mathbb{Z}^{d}$ be initialized to zeros and $b_{R} \in \{0,1\}^d$ denote a binary vector with ones for $R$ and zeros for $R^{c}$. 
\begin{enumerate}
    \item\textbf{Voting}:  repeat  $V^{(i)} = V^{(i-1)} + b_{R_{i}}$   until get $V^{(q)}$;
    \smallskip
    \item\textbf{Cutoff}: given a cutoff parameter $c$, return $\widehat{PA(Y)} = \{j: V^{(q)}_{j} \geq c\}$.
\end{enumerate}

\begin{figure}[h]
\vspace{-1em}
\centering
\includegraphics[width=0.7\linewidth]{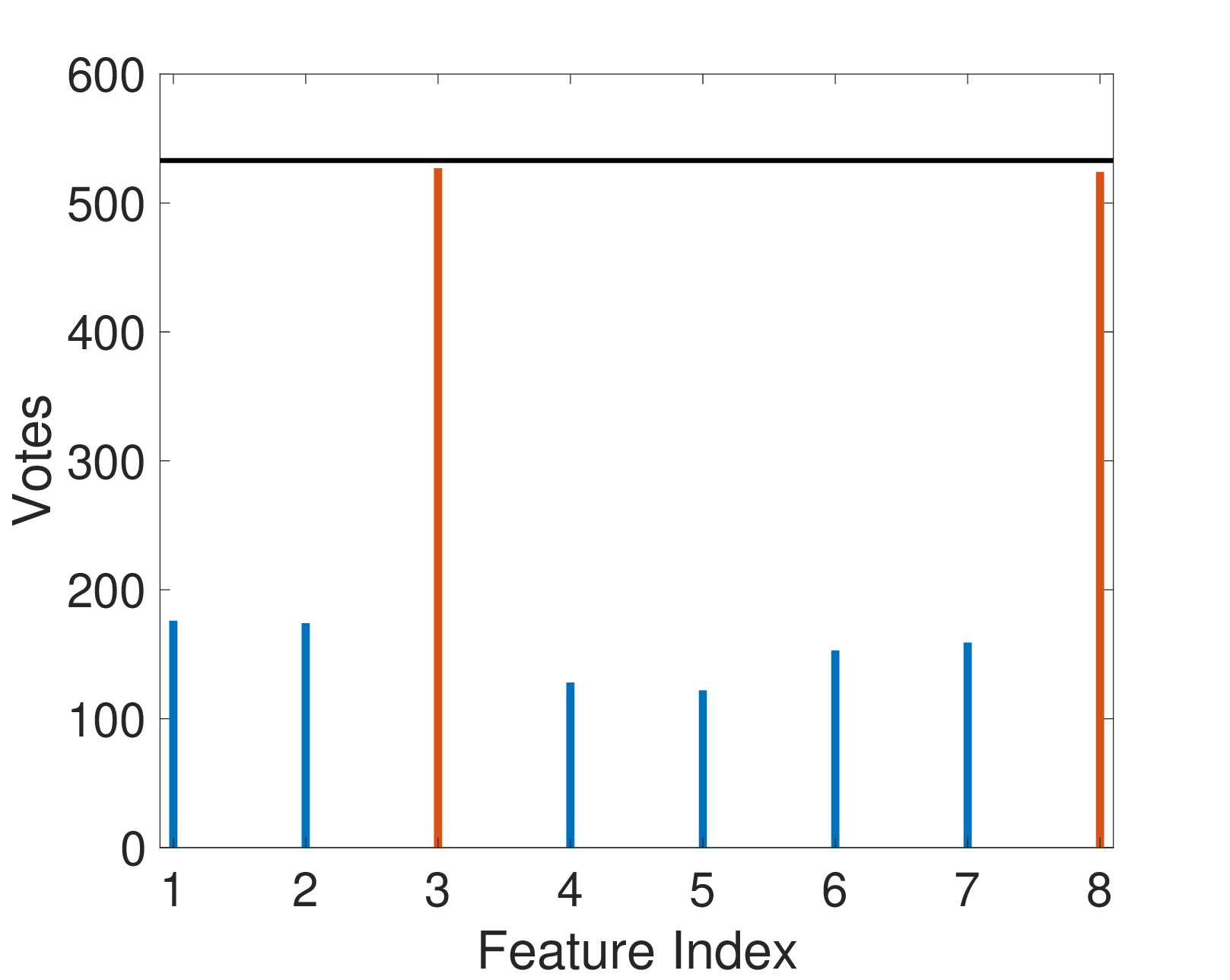}
\caption{An example of the voting results. The black horizontal line represents the total number of IMPs  ($q=533$). Orange lines: the votes for the parents of $Y$; blue lines: the votes for other features. \label{eg_vote}
}
\vspace{-1.5em}
\end{figure}

An example of the voting results is illustrated in Fig.~\ref{eg_vote}, showing that the parents of $Y$ get nearly all the votes but other features receive significantly fewer votes. According to the identifiability results, if $\hat{\mathcal{I}}$ only contains true IMPs, $PA(Y)$ will get full votes, in which case the cutoff parameter is naturally given by $c=q$. However, this will lead to an empty set if $\hat{\mathcal{I}}$ contains at least one non-IMP,  indicating that using $c=q$ will make the procedure too sensitive with respect to non-IMPs. To this end, we propose to use $c = \gamma q$ for some $\gamma \in [0,1]$. Observe that a small $\gamma$ (e.g., $\gamma=0.2$ in Fig.~\ref{eg_vote}) will make $\widehat{PA(Y)}$ include non-causal variables that receive only a small portion of votes. Thus, $\gamma$ controls the algorithm's ability to avoid non-causal variables (or false discoveries).

\section{Synthetic Data Experiments}

\begin{figure*}[t]\centering
\begin{subfigure}{}\includegraphics[scale=0.19]{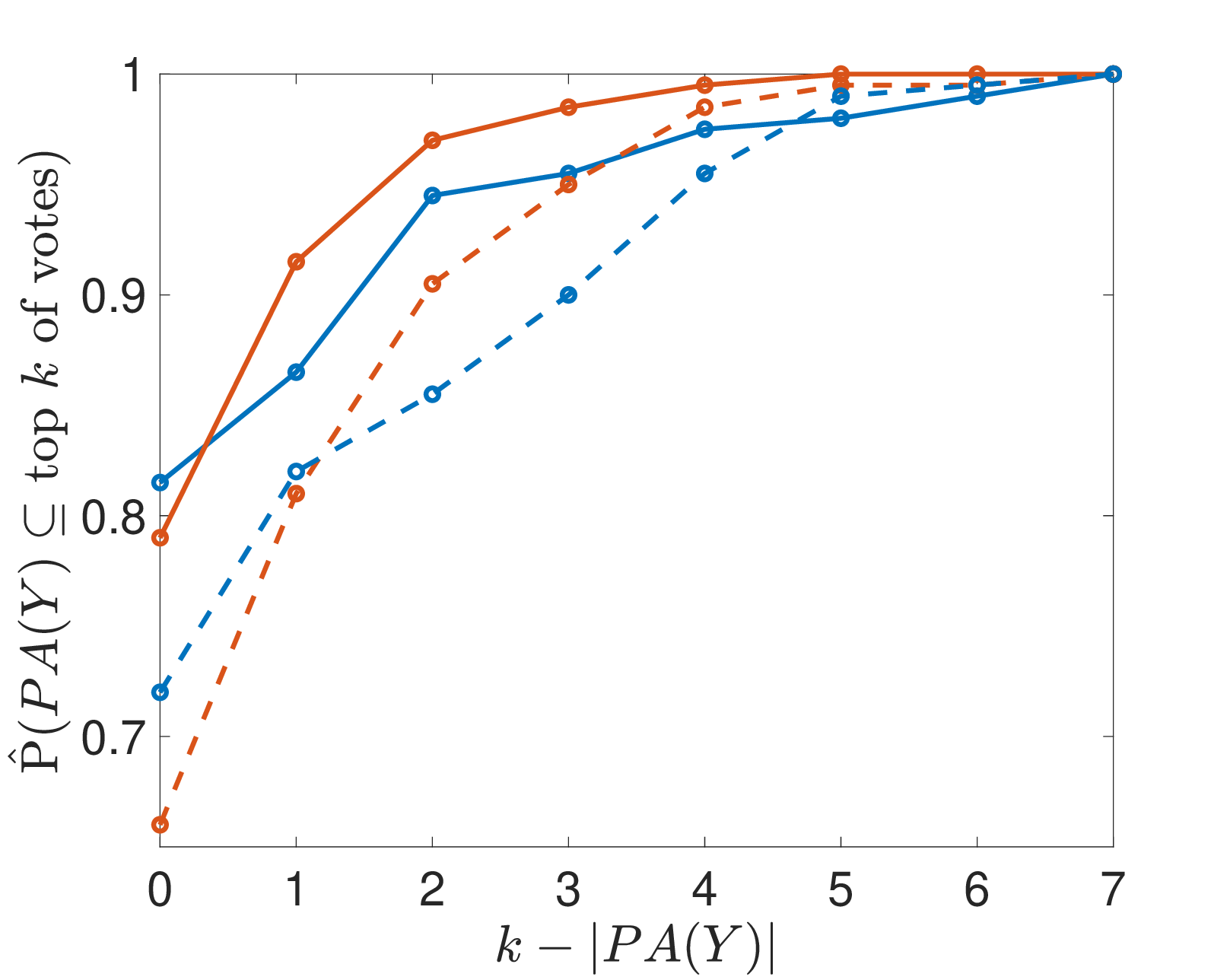}\label{exp6_1}
\end{subfigure}
\begin{subfigure}{}\includegraphics[scale=0.19]{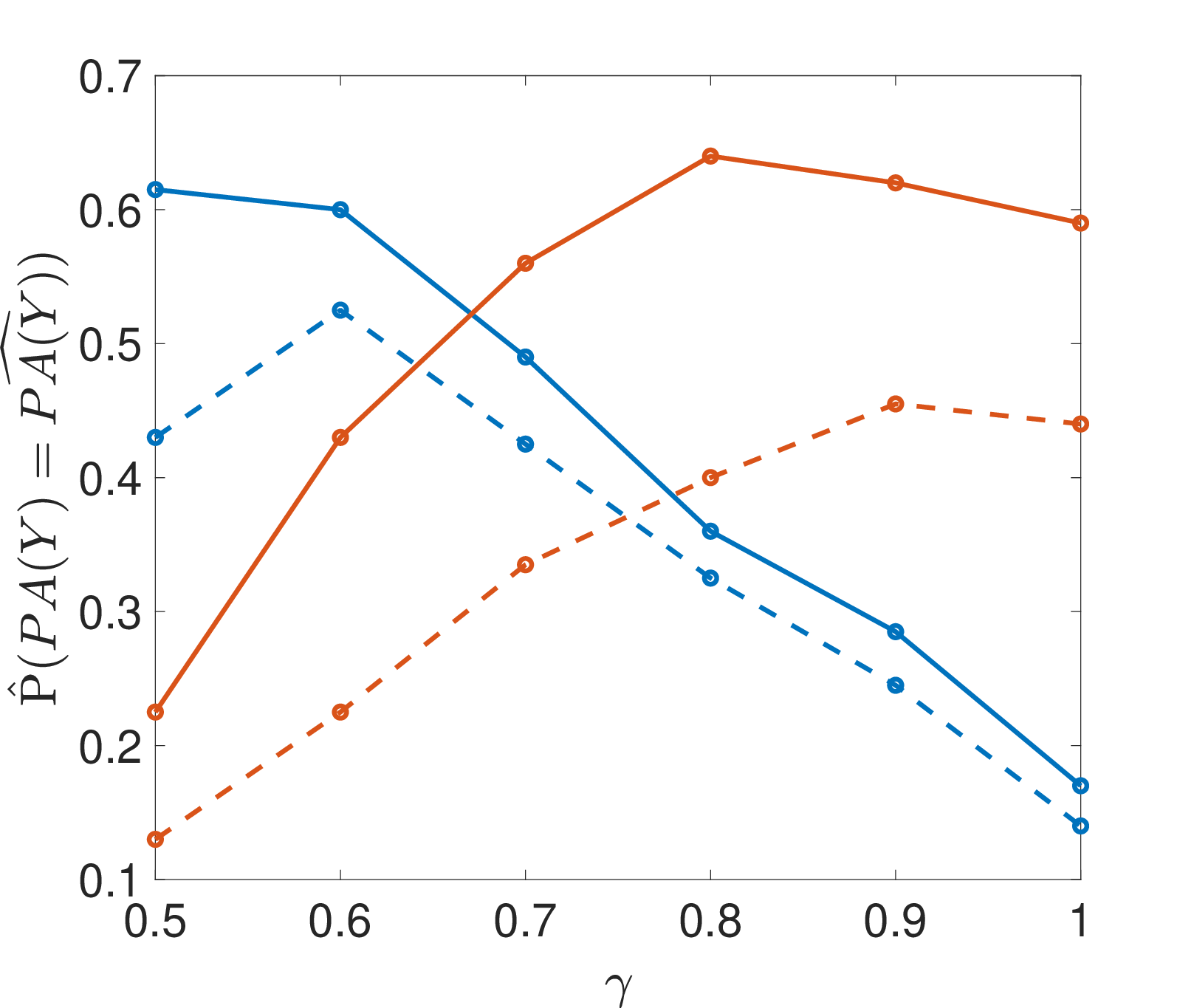}
\end{subfigure}
\begin{subfigure}{}\includegraphics[scale=0.19]{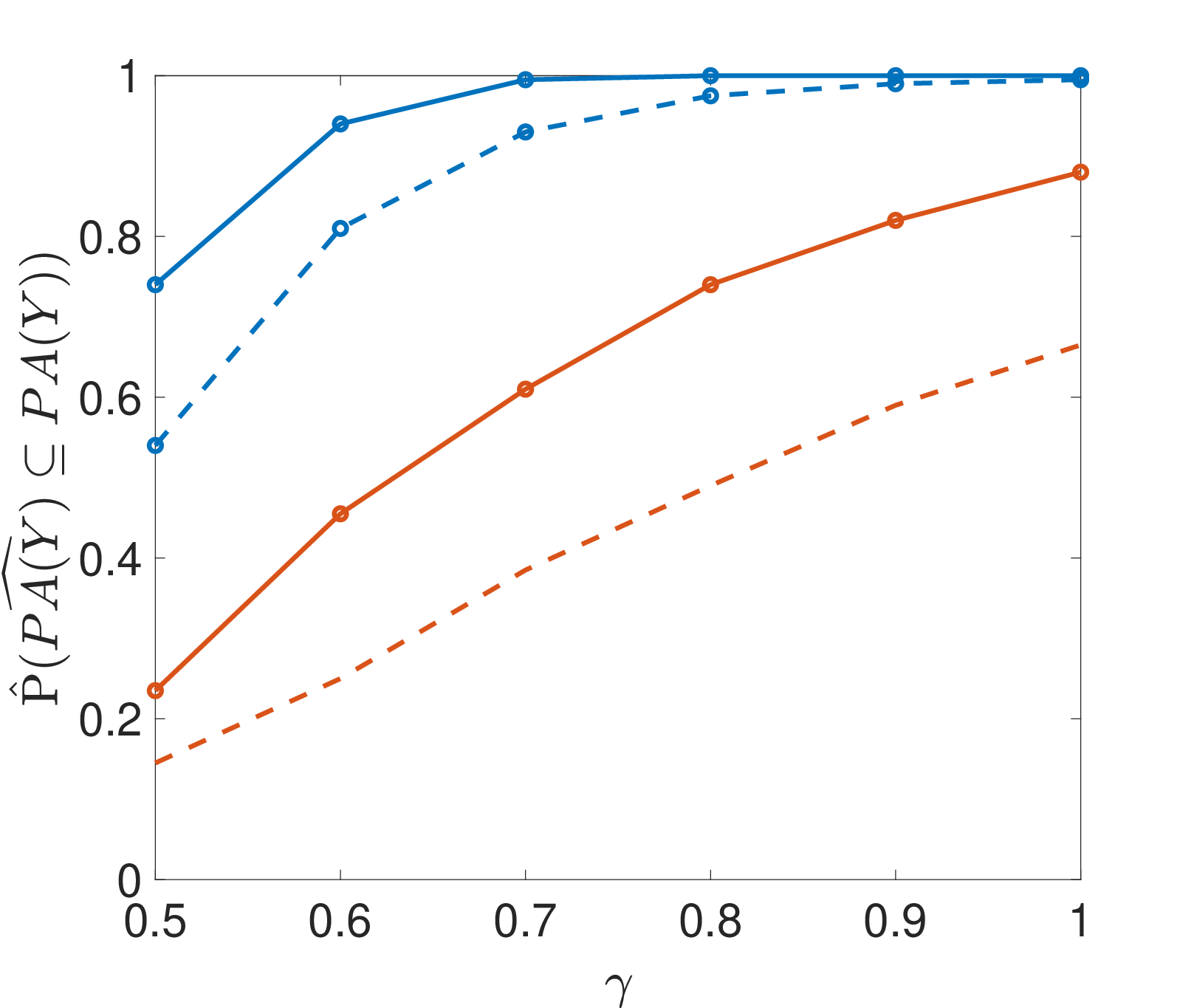}
\end{subfigure}
\caption{\label{our_exp_res}Illustrations of the results of Experiments A (blue lines) and B (orange lines). Solid and dashed lines represent IMP and IMP$_{\text{inv}}$, respectively. From left to right, the first figure shows the empirical probability of the event that $PA(Y)$ is included in the top $k$ of the voting results, as $k$ increases from $|PA(Y)|$; the second figure shows the empirical success probability of the two methods for $\gamma \in \{0.5,0.6,\ldots,1\}$; the last one shows the empirical probability of $\widehat{PA(Y)}$ being a subset of $PA(Y)$.} 
\vspace{-1.5em}
\end{figure*}

We slightly adjust the simulations from~\cite{du2022learning} to demonstrate the strength of IMP and IMP$_{\text{inv}}$ in identifying $PA(Y)$. To make $PA(Y)$ identifiable, we consider the setting when all the parents of $Y$ have changing coefficients across environments. Our main focus is the setting when only $Y$ is intervened, in which case interventional data is more cost-efficient to obtain. Furthermore, we empirically verify whether additional interventions on a randomly selected subset of $X$ are helpful. Our methods are compared with LiNGAM~\cite{shimizu2006linear} and ICP~\cite{peters2016causal}, with the significance levels of all methods fixed at $0.05$. Note that the simulated datasets are generated according to our model assumptions, which could violate the assumptions of the baseline methods. In comparison with ICP, the main advantage of our methods is that the interventional data that (approximately) satisfy our assumptions are much easier to obtain, as discussed in Section~\ref{sec:ICP}. We briefly summarize the experiment settings below. Detailed algorithm descriptions can be found in~\cite[Section~$7$]{du2022learning}. 

We simulate $200$ datasets according to the following setup. First, we generated linear SCMs with random graphs, randomly selected coefficients, and standard normal noise variables. Then, we generate interventional data from $5$ environments in the following two settings, where the sample size is $300$ in each environment. The two settings correspond to Experiments~A.$2$ and A.$3$ from~\cite{du2022learning}, respectively. Since the data from each environment is Gaussian, we apply LiNGAM to the pooled data. 

\noindent\underline{\textbf{A. Intervention on $Y$:}} The response $Y$ is intervened through the coefficients of all its parents and a shift. The intervention on a certain parameter is performed by adding a perturbation term that is sampled from a uniform distribution. 

\noindent\underline{\textbf{B. Interventions on $X$ and $Y$:}} In addition to the interventions on $Y$, there are $4$ randomly selected $X_{j}$'s that are intervened through shifts.

In Fig.~\ref{our_exp_res}, the first figure provides an illustration of the voting results obtained by IMP and IMP$_{\text{inv}}$. When $k= |PA(Y)|$, the estimated probability of $PA(Y)$ being a subset of the top $k$ candidates reduces to the estimated probability of $PA(Y)$ has a higher number of votes than all other features, which is around $80\%$ for IMP (solid lines) and $70\%$ for IMP$_{\text{inv}}$ (dashed lines), respectively; as $k$ increases, the top $k$ candidates will include $PA(Y)$ as a subset with rapidly increasing probability. These observations suggest that $PA(Y)$ generally receives more votes than other features. The last two figures show how the performance of the two methods varies with different choices of $\gamma$. When only $Y$ is intervened (blue lines), IMP exhibits a low percentage of false discoveries for most choices of $\gamma$ (the third figure). In practice, one could choose a cutoff based on the pattern of the voting results since there is often a noticeable gap between the features of the top candidates and others. We demonstrate this idea through two real-world datasets in the next section. Overall, additional interventions on $X$ (orange lines) result in more false discoveries while the success percentage is not significantly higher, indicating that interventions should focus solely on $Y$. The reason behind this is that identifying IMPs is more challenging in Setting~B., resulting in more non-IMPs in $\hat{\mathcal{I}}$.

In contrast, LiNGAM fails to identify $PA(Y)$ in almost all cases ($>99\%$) and always ($>99\%$) identifies a set containing non-causal variables. It is worth noting that even though the pooled data is sampled from a mixture of Gaussian distributions which is non-Gaussian, the model is no longer a linear SCM with independent noise variables, as discussed in Section~7.1 from~\cite{peters2016causal}. While the competitive performance of LiNGAM was reported in~\cite{peters2016causal} when only $X$ is intervened,  LiNGAM appears to be sensitive with respect to interventions on $Y$, as indicated by our experimental results. Similarly, since the key assumption that $Y$ should not be intervened is violated, ICP fails to identify $PA(Y)$ in all cases and the percentages of cases where $\widehat{PA(Y)} \subseteq PA(Y)$ are $30.5\%$ and $43.5\%$ for Experiments A and B, respectively. This shows that, for ICP, the familywise error rate (FWER) of falsely including non-causal variables is not being controlled. However, IMP is able to maintain a low FWER for most choices of $\gamma$ as shown in the third figure of Fig.~\ref{our_exp_res}.

\section{Real-data Experiments}

We consider two real datasets: the first dataset (flow cytometry) is 
interventional data with a well-understood graph structure and known interventions, while the second dataset (COVID) is purely observational data with an unknown graph structure and unknown interventions. The study of the first dataset represents a proof of concept when our assumptions are approximately satisfied. The second one demonstrates the effectiveness of our method, even though our required assumptions are likely to be violated in this complex setting.

\vspace{-1.2em}
 \begin{figure}[h]
\centering
\includegraphics[width=0.66\linewidth]{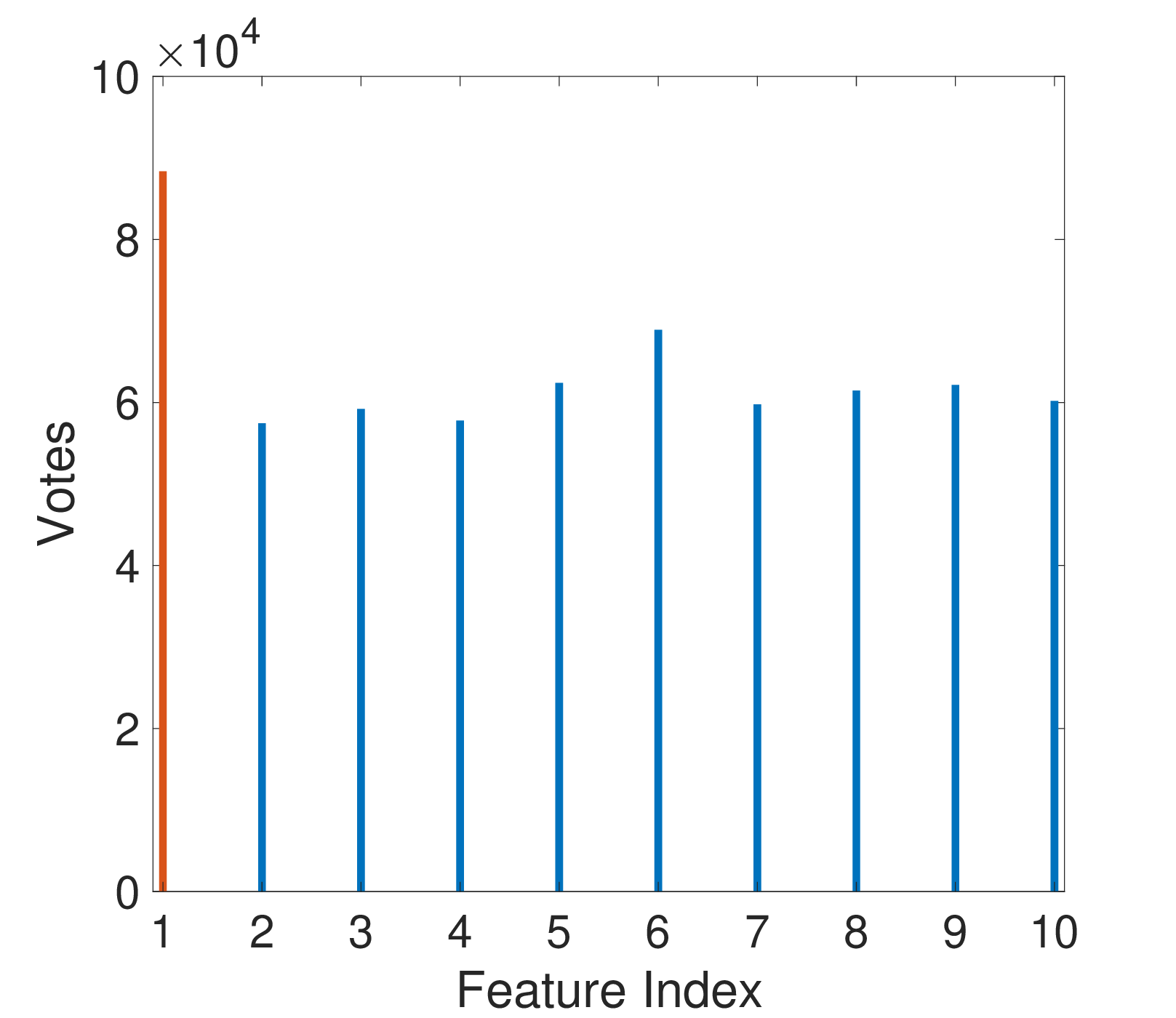}
\caption{Voting results of the flow cytometry dataset. \label{fig_flow}
}
\end{figure}
\vspace{-1em}

 \subsection{Flow Cytometry Dataset}

The flow cytometry dataset~\cite{sachs2005causal} consists of 11 measurements of phosphorylated protein and phospholipid components. There is a well-understood graph structure that describes the interactions among the variables (see Fig 2. from~\cite{sachs2005causal}). The dataset provides interventional data from 14 environments, where different sets of features are intervened at different environments. The sample sizes of different environments range from $700$ to $900$. We conduct the experiments by considering the first $n$ samples from each environment, where $n$ is chosen from
$\{200,300,\ldots,700\}$. There exists one variable called \textsf{MEK$1/2$} that satisfies our assumptions on $Y$, i.e., it has been intervened at least once and it has a child that is not intervened. We consider three environments such that \textsf{MEK$1/2$} is intervened in two and not in the third. In this case, we only present the results of IMP since the voting results of IMP and IMP$_{\text{inv}}$ show a similar pattern. The results of IMP$_{\text{inv}}$ suggest that the invariance property~\eqref{inva_trans} holds approximately for this dataset. From the voting results in Fig~\ref{fig_flow} (with $n=200$), the parent of \textsf{MEK$1/2$}, called \textsf{Raf}, receives the highest number of votes. The black line was not presented in Fig~\ref{fig_flow}, since the highest number of votes only accounts for $46\%$ of the total number of IMPs. The voting results for $n \in \{200,\ldots, 600\}$ all show a consistent pattern as in Fig.~\ref{fig_flow}. But \textsf{Raf} no longer receives the highest number of votes when $n=700$. LiNGAM also selects \textsf{Raf} as the parent of \textsf{MEK$1/2$}; ICP rejects the null hypothesis of the invariance, due to the interventions on the target variable.

\subsection{COVID Dataset}

 The IMP methods have been examined in~\cite{du2022learning} over a COVID dataset~\cite{haratian2021dataset} collected at $3142$ US counties over the time period $01/22/2020 - 06/10/2021$. The target variable of this dataset is the number of COVID cases in a city/county. $5$ major cities from the Western U.S. are chosen as $5$ training environments, which are Los Angeles, San Francisco, San Diego, Seattle, and Phoenix. The dataset consists of 46 predictive features for the number of COVID cases including $12$ temporal features. In~\cite{du2022learning}, the authors considered the time interval $01/03/2020 - 09/30/2020$ ($214$ days) and $10$ temporal features that are not approximately constant in the considered time interval. Both continuous and discrete features are included in the dataset, and the causal relationships among them are unknown, as is often the case in real-world datasets. The IMP method was shown to outperform various baselines for predicting COVID cases in $8$ cities/counties mainly from the East Coast. In this experiment, we aim to understand the estimated IMPs obtained from this dataset from a causal perspective. Specifically, we check whether the features identified by the proposed voting procedure are potential causes for the number of COVID cases.

\begin{figure}[t!]
 \vspace{-.5em}
\centering
\includegraphics[width=0.7\linewidth]{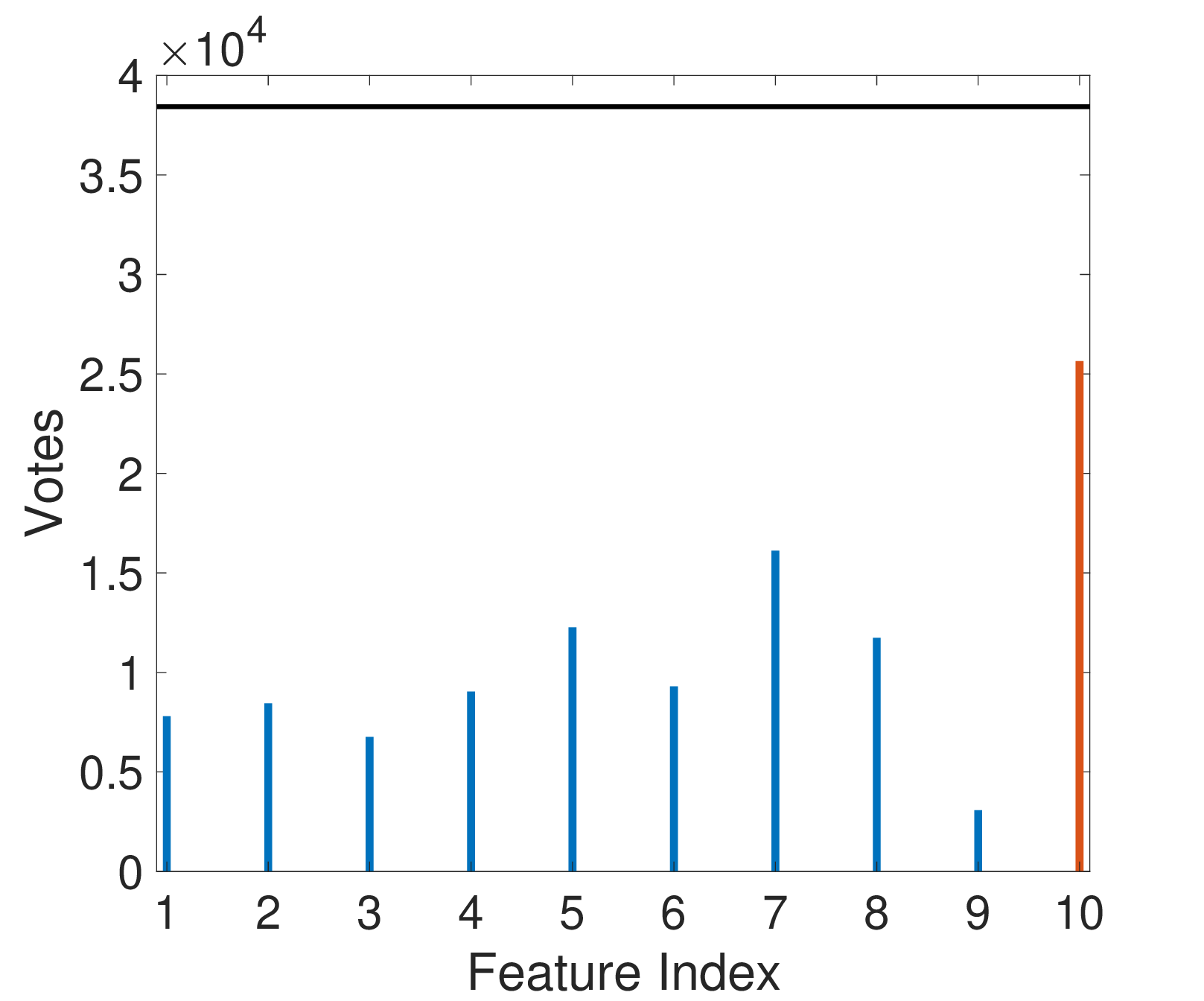}
\caption{Voting results of the COVID dataset.  \label{fig_covid}
}
\vspace{-0.5em}
\end{figure}

 To examine the voting procedure, we focus on the IMP procedure for this dataset since IMP$_\text{inv}$ does not identify any IMPs at a $0.05$ significance level. From Fig.~\ref{fig_covid}, observe that the feature indexed by $10$ receives a significantly higher number of votes than others. This feature, called \textsf{virus pressure}, is defined as the average number of COVID cases in the neighboring counties. Note that \textsf{virus pressure} alone can provide a rough approximation of the number of COVID cases in the county of interest, due to strong correlations among the COVID cases of the neighboring counties. It is thus reasonable to take \textsf{virus pressure} as a key indicator for the number of COVID cases in each county. While LiNGAM identifies $Y$ as a node that is disjoint with all other nodes, which might be because LiNGAM is mainly developed for continuous variables. ICP rejects the null hypothesis that there exists an invariant linear model over $Y$ and any $X_{S}$ at a significance level $0.05$. The results of IMP$_\text{inv}$ and ICP indicate that the invariance assumptions~\eqref{eq:ICP} and~\eqref{inva_trans} are violated, respectively, potentially due to nonlinearities and hidden confounders.

\balance

\bibliographystyle{IEEEtran}
\bibliography{ref.bib}

\end{document}